\documentclass{article}
\usepackage{natbib,latexsym,epsfig,aaspp4}
\def\plotfiddle#1#2#3#4#5#6#7{\centering \leavevmode
\vbox to#2{\rule{0pt}{#2}}
\includegraphics{#1}}

\lefthead{Manning et al.}
\righthead{Primeval Galaxies}

\def\ie{{i.e.,}}

\def\deg{\ifmmode {^{\circ}}\else {$^\circ$}\fi}

\def\secper{\ifmmode \rlap.{^{s}}\else $\rlap{.}{^{s}} $\fi}

\def\kms{\ifmmode {\rm\,km\,s^{-1}}\else
    ${\rm\,km\,s^{-1}}$\fi}
\def\kmsMpc{\ifmmode {\rm\,km\,s^{-1}\,Mpc^{-1}}\else
    ${\rm\,km\,s^{-1}\,Mpc^{-1}}$\fi}
\def\ergAcm2{\ifmmode {\rm\,ergs\,cm^{-2}\,s^{-1}\,{\rm \AA}^{-1}}\else
    ${\rm\,ergs\,cm^{-2}\,s^{-1}\,\AA^{-1}}$\fi}
\def\ergcm2s{\ifmmode {\rm\,ergs\,cm^{-2}\,s^{-1}}\else
    ${\rm\,ergs\,cm^{-2}\,s^{-1}}$\fi}
\def\ergsHz{\ifmmode {\rm\,ergs\,s^{-1}\,Hz^{-1}}\else
    ${\rm\,ergs\,s^{-1}\,Hz^{-1}}$\fi}
\def\ergs{\ifmmode {\rm\,ergs\,s^{-1}}\else
    ${\rm\,ergs\,s^{-1}}$\fi}
\def\ergsA{\ifmmode {\rm\,ergs\,s^{-1}\,\AA^{-1}}\else
    ${\rm\,ergs\,s^{-1}\,\AA^{-1}}$\fi}
\def\ergs{\ifmmode {\rm\,ergs\,s^{-1}}\else
    ${\rm\,ergs\,s^{-1}}$\fi}
\def\WHz{\ifmmode {\rm\,W\,Hz^{-1}}\else
    ${\rm\,W\,Hz^{-1}}$\fi}

\def\spose#1{\hbox to 0pt{#1\hss}}
\def\simlt{\mathrel{\spose{\lower 3pt\hbox{$\mathchar"218$}}
     \raise 2.0pt\hbox{$\mathchar"13C$}}}
\def\simgt{\mathrel{\spose{\lower 3pt\hbox{$\mathchar"218$}}
     \raise 2.0pt\hbox{$\mathchar"13E$}}}

\def\hone{\ion{H}{1}}

\def\heii{\ion{He}{2}$\lambda$1640}
\def\lya{Ly$\alpha$}

\def\nv{\ion{N}{5} $\lambda$1240}
\def\ciii{\ion{C}{3}$\lambda$1909}
\def\civ{\ion{C}{4} $\lambda$1549}

\begin{document}
\title{The Dust Temperature of the ``Dusty'' Radio Galaxy MG
1019+0535: \\
Evidence for an Outflow}
\author{Curtis Manning \& Hyron Spinrad}
\affil{Department of Astronomy, University of California at Berkeley \\
Berkeley, CA 94720 \\
{\tt email: (cmanning,spinrad)@bigz.berkeley.edu}}

\begin{abstract}

Radio galaxies characteristically have strong \lya\ emission lines.
However, a few have \lya\ equivalent widths that are substantially
weaker in relation to other emission lines.  One in particular was
studied by \citet{Dey:95}, MG 1019+0535 ($z=2.765$).  We report on our
reduction of Infrared Space Observatory (\emph{ISO}) data in the 160
$\mu$m-band for this galaxy.  We also compile information on two other
high redshift active galaxies with weak \lya\ lines, the radio galaxy
TXS 0211--122, and the AGN--starburst galaxy F 10214+4724, to
provide a small weak-\lya\ line sample.  IRAS plus ISO data show that
F 10214+4724 has a temperature $89 \pm 12$ K.  TXS 0211--122 was not
detected in either the submillimeter or microwave.  Submillimeter
measurements of MG 1019+0535 \citep{Cimatti:98} were suggestive of a
dust temperature in the range $35 \,{\rm K} \leq T_d \leq 80$ K.
However our 2-$\sigma$ upper limit on the flux at 160 $\mu$m shows
that $T_d \lesssim 32 \, {\rm K}$.  An energy argument based on
observations which constrain the total optical extinction strongly
suggests that the dust temperature must be even lower; $T_d \lesssim
20$ K.  We find the contrast between the high-temperature dust in the
active starburst galaxy and the low-temperature dust in the evolved,
albeit active galaxy, is consistent with an expanding cloud of dust.
We find that the temperature range can be reconciled if we are seeing
MG 1019+0535 at a post-starburst age of $\approx 500-700$ Myr,
with the bulk of its dust cloud at a galactocentric distance $R_d
\gtrsim 300$ kpc.

\end{abstract}



\section{Introduction}

Though the role of dusty \hone\ clouds in the quenching of the \lya\
line of star forming galaxies is by now well-recognized
\citep{Neufeld:91,Chen:94,Legrand:97,Kunth:98b}, how it is manifested
in specific galaxies is not well-understood.  Real progress on the
observational side of the problem began with the launch of the
\emph{Infrared Astronomy Satellite} (IRAS).  With its detectors tuned
to 12, 25, 60 and 100 $\mu$m, it was sensitive to dust emission at
moderate temperatures from galaxies in the low-redshift universe.  The
dust emission at these wavelengths may be used to determine the
spatial extent of dust and help constrain the total extinction
occurring in the galaxy.  The often tantalizing data emerging from
IRAS led to the planning and implementation of more sensitive
detectors, both ground- and space-based. The Infrared Space
Observatory (ISO)\footnote{ISO is an ESA project with instruments
funded by ESA Member States (especially the PI countries: France,
Germany, the Netherlands and the United Kingdom) with the
participation of ISAS and NASA.}, was designed to give higher spatial
and spectral resolution and greater sensitivity over this far-infrared
(FIR) range.  Astronomy now appears to be approaching a ``golden age''
of FIR observation, as telescopes, observatories, and instruments,
such as SCUBA, MIPS on SIRTF, and SPIRE on FIRST, are implemented.

How is dust produced, and what is its fate?  A very simplified model
of dust production in galaxies starts with gasses expelled from
evolved stars --- supernovae and winds from evolved, metal-enriched
stars.  This enriched gas forms dust when condensing molecules collide
and adhere.  The probability for survival of a small grain is largely
dependent on the number density and energy of ambient UV photons; the
absorption of a single UV photon can vaporize a small grain.  A burst
of star formation may cause large-scale winds which carry dust to
large galactocentric distances
\citep{Legrand:97,Frye:98,Warren:98,Kunth:98b}.  In addition, the
expelled dust may be propelled by photon pressure and corpuscular drag
\citep{Burns:79}.  The increased incidence of FIR/sub-mm emission from
dust in radio galaxies with increased redshift \citep{Archibald:00}
suggests that these galaxies were rich in warm dust shortly after the
star-forming episode which formed them, but also that in time, the
dust is dispersed.  This implies that the efficiency of the galactic
dust in absorbing optical radiation declines strongly with time
following a starburst, leading to a declining color excess.

One study of IRAS fluxes in galaxy disks \citep{Persson:87}
found two distinct dust components; a warm component associated with
OB associations and HII regions, and a cool component associated with
a neutral ISM.  Similarly, recent studies \citep{Calzetti:99,
Calzetti:00} find that dust in nearby star-forming
galaxies can often be fit by two-components -- a warm
(typically $T\approx 40-55$ K), and a cool ($T\approx 20-23$ K)
component.  A wavelength dependence of the dust emissivity ($\propto
\sim \nu^{\epsilon}$) produces a modified Planck function with a
luminosity proportional to $T_d^{4+\epsilon}$, where $\epsilon$ is the
dust emissivity index, usually taken to be in the range $1\leq
\epsilon \leq 2$.  Galactic dust has an emissivity index that
increases slowly from the Rayleigh-Jeans tail, toward a value $\sim
\nu^2$ near the maximum, and is well-fit by an emissivity $\sim
\nu^{1.7}$ with a median dust temperature $T_d \simeq 19$ K
\citep{Finkbeiner:99}.  However, the fits to local star-forming
galaxies by \citet{Calzetti:00} is well-approximated by an $\epsilon
\sim 2$.  Calculations of dust mass in low redshift, actively
star-forming galaxies \citep{Calzetti:00}, find that the cool dust
mass may often be more than two orders of magnitude larger than the
warm component.  Yet because of the strong wavelength dependence of
the emission, the warm component is usually more luminous.  A high
dust opacity has been cited as a factor in the production of cool
clouds \citep{Calzetti:00}, though in the absence of a warm component
it may be reasonable to conjecture that the low dust temperature is
due to its having been blown to large galactocentric distances in a
post-starburst galaxy.

We now apply these concepts to radio galaxies.  Radio galaxies are
often viewed as a sub-population of elliptical galaxies
\citep{Matthews:64, McLure:99}.  As a group, ellipticals show few
signs of significant amounts of dust in the lower-redshift universe.
But at high-redshift, closer to the epoch of galaxy formation, it may
be possible to see the signs of large amounts of dust in radio
galaxies.  One method of detecting this dust is to observe the
strength of its \lya\ emission line relative to other emission lines.
In their study of the radio galaxy MG 1019+0535 (z=2.765),
\citet{Dey:95} discussed the implications of the observed low flux in
the \lya\ line in relation to that of \heii\, and pondered the role of
dust in the apparent selective extinction of its \lya\ line.  They
found that relative to the \heii\ line, the observed \lya\ line
strength is only $\sim 10$\% relative to that of the composite 3CR
radio galaxy spectrum of \citet{McCarthy:93}.  \citet{Dey:95} also
discussed the radio galaxy TXS 0211--122 ($z= 2.340$), and IRAS galaxy
(and faint radio source) F 10214+4714 ($z = 2.286$), both of which
have weak \lya\ to \heii\ ratios $< 2.5\%$, and 15\%, respectively.

Because the \lya\ line center of MG 1019+0535 is redshifted relative
to other permitted lines, \citet{Dey:95} proposed a scenario in which
the dust cloud lies outside the line-emitting region, perhaps
propelled away by winds.  Dust has a well-modeled, broad effect on
optical spectra, producing a spectral reddening by its
wavelength-dependent extinction of photons.  However, if neutral
hydrogen is associated with the dust cloud, then \lya\ photons will
experience a much greater extinction due to their increased
path-length in the cloud arising from their resonance with \hone.

In an attempt to detect emission from the postulated dust cloud, we
observed MG 1019+0535 with the Infrared Space Observatory (ISO).
Using this, and sub-millimeter data from the literature, we
investigate its dust content and temperature in relation to measures
of its optical extinction.  We also retrieved archival ISO data on TXS
0211--122 and F 10214+4724 which we use to compare with our results on
MG 1019+4724.

The paper is organized as follows: In \S2 we present the ISO data for
the three galaxies.  In \S3 we describe the reduction of {ISO} data,
and introduce other, submillimeter and IRAS data in order to derive
dust temperatures.  In \S4 we attempt an energetic reconciliation of
the FIR, with the optical data.  In \S5 and \S6 we discuss our
findings, and present our conclusions.

Throughout this paper we assume an Einstein-De Sitter cosmology with
$H_0=50~\kmsMpc$, unless otherwise noted.

\section{ISO Data}
\subsection{MG 1019+0535}

Our observations of MG 1019+5035 were made with ISOPHOT, the imaging
photopolarimeter on-board ISO, in observing mode PHT 22.  ISOPHOT
offers multi-filter photometry in two mosaic arrays, equipped with
filters in wavelength bands ranging from 50 to 100 microns (the C100
detector, a 3 by 3 pixel grid), and 120 to 200 microns (the 2 by 2
pixel C200 detector).  Preceding and following each measurement of the
galaxy, observations of an internal Fine Calibration Source (FCS) are
made.  The FCS measurements are designed to act as a stable reference
source to be used for calibration of the data. Experiments are
generally designed so that the standard source is heated to a
temperature which will give a flux at the detectors approximately
equal to that expected from the astrophysical source plus background.
This is because there is a large temporal lag in the sensitivity
stabilization for large changes in flux.  Observations were made
through the 50, 90, and 160 $\mu$m-band filters.  For reasons
explained below, we only use the 160 $\mu$m data, a filter which is
centered at 174 $\mu$m.  Two measurements were made, a $256~{\rm s}$
exposure, then a second exposure, spatially offset from each other by
$90^{\prime \prime}$ in declination, for $128~{\rm s}$.

As we shall see, analysis of the images shows that the calibration of
the data is not satisfactory, perhaps due to a non-linearity in the
response of the detectors to different flux levels.  One result,
clearly seen in the $160 ~\mu$m data, is a wide variance among the
individual pixels of these observations whose pixel-specific values do
not significantly change when the telescope pointing is changed.
Unfortunately these systematic variations are not constant in time,
and so cannot be modeled by recourse to other data except by
considering observations which are in close temporal proximity.
However, only the $160 ~\mu$m data had different pointings, so that
the 60 and 90 $\mu$m data cannot be corrected for these problems.

\subsection{TXS 0211--122 and F 10214+4724}
ISO data for TXS 0211--122 and F 10214+4724 were available from the
archives (http://ines.vilspa.esa.es/ or
http://www.ipac.caltech.edu/iso/).  These data were taken in P32 mode,
which produces a chopped map.  We discuss this more thoroughly in
\S \S3.4 and 3.5.

\section{Data Reduction}
\subsection{MG 1019+4724}

We will describe the reduction of the ISO data for MG 1019+0535 in
detail, as it is unconventional.  As noted above, we consider only the
160 $\mu$m C200 data.  ISOPHOT data is reduced using standard PHT
Interactive Analysis (PIA) software, a software package developed
specifically for the reduction of ISOPHOT data.  To distinguish
calibrator measurements from those of the astrophysical object, we
refer to the former as \emph{FCS}, and the latter as \emph{source}
measurements.

\begin{figure}[hhh] 

\plotfiddle{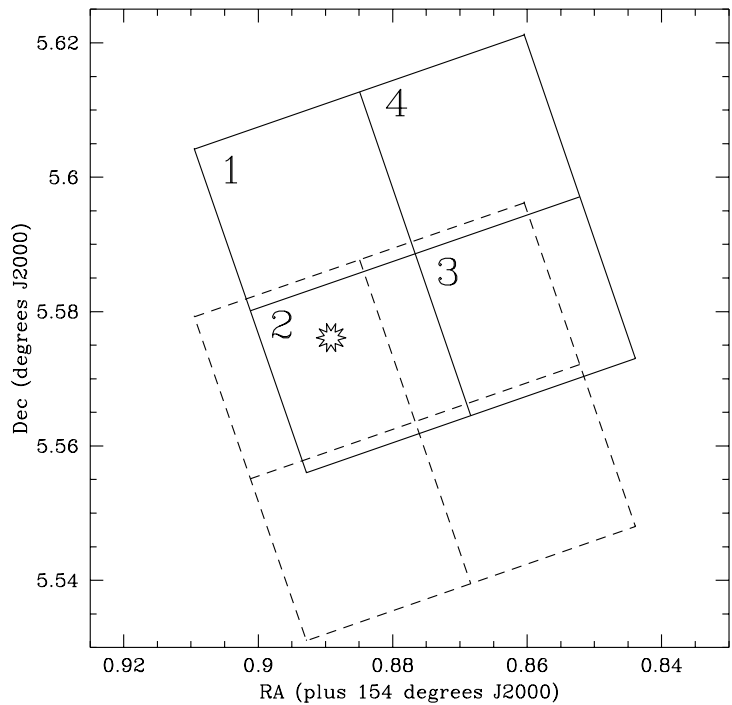}{3.7in}{0}{50}{50}{-146}{-75}

\caption{The placement of the C200 detectors in the sky relative to MG
1019+0535, symbolized by the star.  The radio galaxy is reasonably
close to the center of the pixel in which it resides.  Numbers show
pixel numbers, referred to in the text.  Each pixel is about
$90^{\prime \prime}$ square; the airy disk in the 160 $\mu$m-band
filter has a radius of $\sim 67^{\prime \prime}$.}
\end{figure}

Figure 1 shows the placement of the detectors on the sky during the
two source measurements.  As can be seen, MG 1019+0535 was in pixel
\#2 during the first measurement, and pixel \#1 in the second.  We
expect that any significant flux density in the 160 $\mu$m-band from
the radio galaxy would be betrayed by a different pattern in the two
measurements.  We reduced the data to the point just before the flux
calibration, then studied both the FCS and source data.  We noticed
that specific pixel values were very consistent between successive
observations, but among themselves, pixel values varied by as much as
$\sim 40\%$.  The two source observations of MG 1019+0535 also show a
wide variation in values among pixels of a given measurement, and
which are also highly consistent between measurements.  However, the
pixel ranking is different between FCS, and source measurements.  The
mean pixel values of the source measurements of observations \#1 and
\#2 are different by $\sim 4\%$.  This is in all probability
attributable to an internal systematic effect due to the different
length of the observations (256, and 128 s, respectively). By
normalizing the means of the source, and FCS measurements between
observations \#1 and \#2, it is found that the average variation
within the individual pixels in the two measurements is reduced to on
order $1.2\%$, and $1.4 \%$ of the mean normalized source, and FCS
measurements, respectively.

Three preliminary conclusions can be drawn from this data.  First,
there is a clear consistency between the sets of FCS and source
measurements.  Second, taken together with the large variance of
values between pixels of both source and FCS measurements, this
suggests either a severe non-linearity in the response to an energy
source, variations in detector response between pixels, or additional
sources of systematic error.  Finally, the very small variation of
measurements of the source as a function of pixel number, taken
together with the movement of the telescope (see Fig. 1) implies that
the source radio galaxy does not make a significant contribution to
the flux detected in the pixels.  Therefore, a good first
approximation would appear to be that there is no sign of a signal
from the source galaxy.  Subsequent manipulation of the data appears
to confirm this conclusion.

\subsection{Procedures for Further Calibration of MG 1019+0535 Data}

In order to test for the existence of a signal of the radio galaxy in
the data, we formed the null hypothesis that there was no signal.
Since geometric factors are involved in the calibration process, it
was thought wise to use the calibrated data (which uses the
sensitivity function derived from the FCS measurements) as a basis for
further reductions.  The fact that the radio galaxy is in pixel 2 in
the first, and in pixel 1 in the second measurement, should enable our
hypothesis to be invalidated.

Table 1 shows the sequence of calibration.  Step 1 shows the fluxes in
MJy ${\rm sr}^{-1}$ following calibration of the data.  In step 2, we
normalize the flux densities of observation \#2 to those of
observation \#1 by multiplying the pixel values in the former by the
ratio of the mean of the pixels in the first source measurement to
that in the second, a reduction of $\sim 4\%$.  We normalize
measurement \#2 to \#1, rather than the reverse because \#1 has a
twice longer exposure, and is thus more likely to be correct.  The
third step corrects for the apparent variation in sensitivity of each
pixel; each pixel count in step 2 is multiplied by the average of all
pixels in step 2 divided by the pixel-specific average between
measurements.  Finally, in step 4 we subtract off the mean.  Note that
step 3 has the effect of making the deviations from zero symmetric
between measurements 1 and 2. We discuss this more extensively below.
In Table 1, the data are presented in units of MJy/sr.  The asterisks
denote the presence of the radio galaxy in that pixel.

\begin{center}
\begin{tabular}{c|cc||cc||cc||cc} \hline

 & Step  One &  & Step  Two &  & Step  Three &  & Step  Four \\
\hline
Pixel No. &  Meas. 1 &  Meas. 2 &  Meas. 1 &  Meas. 2 &  Meas. 1 &  Meas. 2  &  Meas. 1 &  Meas. 2\\
\hline
1  & 7.66  & 7.93* & 7.66  & 7.627* & 5.888  & 5.862* &  0.013  &  -0.013*   \\
2  & 5.38* & 5.55  & 5.38* & 5.341  & 5.891* & 5.895  &  0.016* &  -0.016   \\ 
3  & 6.12  & 6.49  & 6.12  & 6.234  & 5.826  & 5.925  & -0.049  &   0.049   \\ 
4  & 4.34  & 4.47  & 4.34  & 4.298  & 5.903  & 5.847  &  0.028  &  -0.028   \\
\hline
\end{tabular}
\end{center}

Table 1 : The reduction of data in units of ${\rm MJy} \, {\rm
sr}^{-1}$.  See the text for a discussion of the steps.  The asterisk
denotes the pixel in which MG 1019+0535 was located during the
measurement.

Referring to the final step of Table 1, one can see that if there was a
significant signal from MG 1019+0535, then the variance in pixels 1
and 2 would probably be larger than that in pixels 3 and 4.  Also
pixel 1 is expected to have an excess in measurement 2, and pixel 2 is
expected to have an excess of measurement 1.  Of these three
conditions, only the very last is satisfied, but only weakly so,
because pixel 4 has a higher reduced flux density in measurement 1
than has pixel 2.  Thus it appears that the source is not detected in
the $160 ~\mu$m-band.

It is possible to derive an upper limit to the flux density by an
analysis of these data, assuming the median flux densities are
reasonably close to reality.  The above data has a standard deviation
of $\sigma=0.016 \,{\rm MJy} \, {\rm sr}^{-1}$.  However, it is
probable that that the true background-subtracted pixel flux densities
do not lie equidistant between the two measured background-subtracted
and calibrated measurements in each pixel, as we have constructed it
to be.  That is, if our null hypothesis is correct, and there is no
significant signal from MG 1019+0535, then the background-subtracted
and calibrated measurements in each pixel are samples from a
distribution of unknown mean, but which have been adjusted to give an
artificial mean equidistant from the two measurements.  A simulation
showed that the intrinsic standard deviation is $\sigma = 0.022$.
Thus the 1-$\sigma$ upper limit on the existence of a signal from the
source in a given pixel is 0.022 ${\rm MJy} \,{\rm sr}^{-1}$, or 4.2
mJy per pixel.  This would be the $1-\sigma$ upper limit of the source
if its whole flux landed in one pixel.  Because of the large airy disk
at 160 $\mu$m, if the source is in the center of a C200 pixel, then
approximately 64\% of the total flux at 160 $\mu$m lands in that pixel
\citep{Laureijs:99}.  This correction raises the 1-sigma constraint is
$\sim 6.6$ mJy.  The 95\% confidence upper limit on the flux density
from MG 1019+0535 at 160 $\mu m$ is thus $F_{160} \approx 13.2$ mJy.


\subsection{Submillimeter data for MG 1019+0535}

Recent observations of AGN using IRAM at 1250 $\mu$m and JCMT (the
James Clerk Maxwell Telescope, Mauna Kea, Hawaii) at 790 $\mu$m
\citep{Cimatti:98} provided the first positive measurements of dust
associated with the radio galaxy MG 1019+0535.  They find the flux at
790 $\mu$m is $14.70 \pm 4.60$ mJy, and at 1250 $\mu$m it is $2.13 \pm
0.47$ mJy.  Published observations of synchrotron emission at longer
wavelengths were cited that indicated that the possibility of
contamination of their sub-millimeter observations was small and
negligible in the observed 1250 and 790 $\mu$m wave-bands,
respectively.  The flux densities in these wave-bands confirmed that
they were sampling the Rayleigh-Jeans part of the thermal
dust-spectrum.  Though these data do not enable a resolution of the
temperature, \citet{Cimatti:98} thought it likely to be within the
range $35 \leq T_d \leq 80 \, K$, for the optically thin, and
optically thick cases, respectively.




Using the sub-mm data of \citet{Cimatti:98}, together with our
estimated 2-$\sigma$ upper-limit flux density of 13.2 mJy at an
observed 160 $\mu$m, we search for a least-squares fit to a Planck
spectrum modified by the assumed dust emissivity index $\epsilon = 2$.
The resulting least-squares fit is consistent with a large FIR
luminosity ${\cal L}_{FIR} \simeq 3.10 \times 10^{46} ~\ergs$, and an
upper limit dust temperature of $T \sim 32.2$ K, which is below the
range of the warm component of dust emission from star-forming
galaxies \citep{Calzetti:00}.  The 1-$\sigma$ upper limit (\ie\ 6.6
mJy) is fit by $T_d \leq 29.4$ K with a FIR luminosity ${\cal L}_{FIR}
= 2.25 \times 10^{46} ~\ergs$.

Due to the extraordinary measures required to reduce the ISO data, it
was thought prudent to check the reliability of our findings.  In
addition to serving this purpose, the following analysis of data for F
10214+4724, viewed as an AGN $+$ starburst, may provide a glimpse of
how MG 1019+0535 may have appeared at an earlier time.

\subsection{ISO and IRAS data for F 10214+4724}

The ultraluminous IRAS galaxy F 10214+4724 is a weak radio source with
a Seyfert 2 nucleus and a circumnuclear starburst
\citep{RowanRobinson:93,Goodrich:96,Lacy:98,Serjeant:98}.  It has been
shown to be lensed \citep{Graham:95,Broadhurst:95,Downes:95}, but
remains one of the most intrinsically luminous galaxies known.
\citet{Dey:95} noted that the \lya\ line was extinguished by a factor
greater than 40 relative to the \heii\ line, though
\citet{Serjeant:98} finds a double-peaked \lya\ line with a total flux
nearly equal to that of \heii , leading to an extinction ratio of only
$\sim 10$.  Due to its non-spherically symmetric Seyfert 2 nature, we
do not expect the \lya\ to \heii\ ratios to be as significant as with
the radio galaxy MG 1019+0535 with respect to issues of dust
distribution.

The ISO data for F 10214+4724 was taken in observing mode PHT32 by
U. Klass (observation \#14500967 and \#14500968).  This data has
similar calibration problems to that of the PHT22 data, however, the
over-sampling of this observing mode has resulted in a partial
smoothing of the effects.  Nevertheless there are still large
systematic effects apparent on the maps, which take the form of
patchy, systematically high or low values near the beginning or end of
the raster mapping process.  These spatially distinct fluctuations
must have been caused by temporal fluctuations of sensitivity in this
$3 \times 1$ raster map.  The data is over-sampled in the ``x''
direction by the chopping, however it has only two pixels in the ``y''
direction.  Figure 2 shows schematically the reduced 180, and 200
$\mu$m data (left, and right sides of the figure, respectively) along
these two lines of pixels.  The galaxy is located on the line
separating these two lines of pixels, and should have an equal
contribution to each.  The difference in the contributions of the
upper and lower pair of pixels apparent in the upper pair of
illustrations is attributable to calibration systematics between the
pixel pairs.  By carefully locating the source on the maps, we were
able to mask out the area dominated by our source.  We then fit a
fourth-order polynomial to the remaining data, which was then
subtracted from the data.  The integrated flux density for the 180 and
200 $\mu$m dat was found to be 198, and 82 mJy, respectively.  The
large diameter of the airy disk for 180 and 200 $\mu$m (151 and
$168^{\prime \prime}$, respectively) may have resulted in a
significant fraction of the signal falling within the data used for
fitting the polynomial model.  For the 180, and 200 $\mu$m
measurements, the airy disk extends to 5.9 and 7.0 sub-pixel units
(see Fig. 2) to either side of the peak of the emission, respectively,
whereas only a total of 8 pixels were blocked out in the fitting of
the continuum.  Because this process could only over-estimate the
background, our reduced fluxes should be considered lower bounds on
the flux densities.

The IRAS data for F 10214+4724 shows a flux density of $204 \pm 46$,
and $580 \pm 150$ mJy in the 60 and 100 $\mu m$--bands, respectively,
as retrieved on NED (Jan., 2000).  By transforming to $F_{\lambda}$,
we find almost identical fluxes ($F_{\lambda} \approx 1.70 \pm 0.40$,
and $1.76 \pm 0.44 \times 10^{-17} ~\ergsA$, respectively).  We fit
these values to a Planck spectrum with an emissivity index
$\epsilon=2$ by placing them on opposing sides of the maximum, and
using the spectral responsivities of the 60 and 100 $\mu$m filters
(available from the Goddard Space Flight Center web-site), we derive a
normalization and temperature which are consistent with the observed
fluxes.  We find $T_d \simeq 89 \pm 12$ K, and integrated from 1 to
500 $\mu$m, the attributed FIR luminosity is ${\cal L}_{FIR} \simeq
7.4 \pm 0.8\times 10^{47} ~\ergs$.  However, observations at the JCMT
\citep{RowanRobinson:93} at 450 and 800 $\mu$m, show this model is an
order of magnitude low in $F_{\nu}$.  This excess over the model
suggests a massive, but cooler cloud ($T_d \approx 30$ K).

Analysis indicates that the FIR luminosity is magnified by $\sim 13$
times \citep{Downes:95}.  Neglecting the luminosity contributed by the
cooler cloud, this leads to a more modest FIR luminosity ${\cal
L}_{FIR} \approx 5.7 \pm 0.6 \times 10^{46} ~\ergs$.  When the fit is
extended to the 180 and 200 $\mu$m bands we find the ISO data is $\sim
55\%$, and $\sim 27\%$ lower than that predicted by the fit,
respectively, for the central value of the temperature, but they are
only modestly low when the higher side of the $1-\sigma$
temperature range is used.  It is probable that the ISO data appears
low because the background continuum level was over-estimated, as
noted above.  Thus, with these caveats, the ISO fluxes appear
consistent with IRAS fluxes, at least to within a factor of two.  We
conclude that the upper limits on $F_{160}$ for MG 1019+0535 (\S3.2)
can be accepted with confidence.
  
\begin{figure}[hhh] 
\plotfiddle{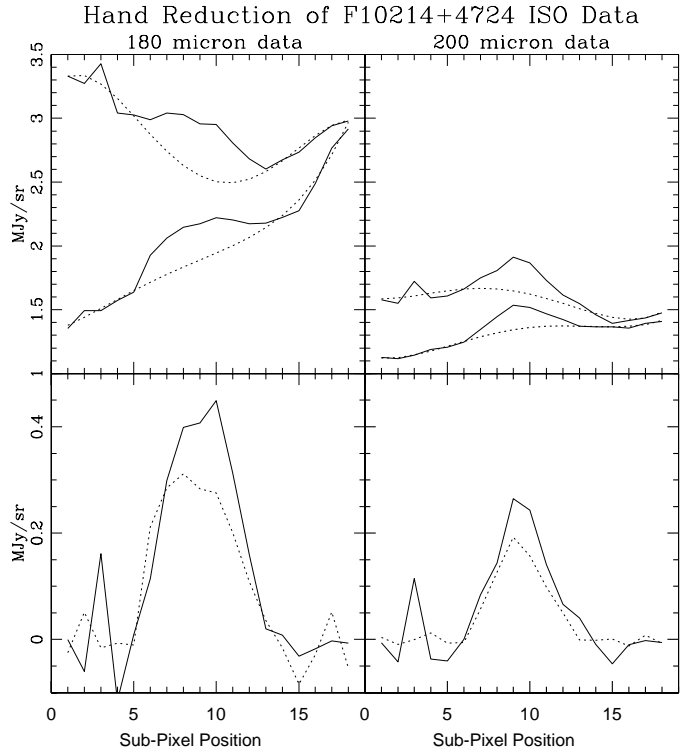}{3.7in}{0}{50}{50}{-146}{-75}
\caption{The flux-calibrated emission in the two lines of pixels from
the $3 \times 1$ raster map of F 10214+4724 ISO data in 180 and 200
$\mu$m (upper left and right, respectively).  In the upper plots, the
solid line represents the data, and the dotted lines represent the
4-th order polynomial fit made by considering the points between
pixels 1 through 5 and 14 and 18.  In both cases the summed flux from
the top row of pixels (numbers 1 and 4 in Fig. 1) have the higher
counts.  The bottom two panels show the fit subtracted from the data,
where the solid lines represent the top row of pixels, and the dotted
lines represent the contributions of pixels 2 and 3.  This systematic
difference again hints at unremoved systematic errors.  One can sense
that the extracted fluxes are lower limits, as the wings of the PSF
were included in the data for the continuum fit (see text).}

\end{figure}


\subsection{ISO and Sub-millimeter Data for TXS 0211--122}

Because the modestly lower redshift of TXS 0211--122 ($z_{em} = 2.300$)
would place the 160 $\mu$m band closer to the maximum on the Wien side
if its dust cloud had a temperature close to that of MG 1019+0535, it
was hoped that a detection was possible.  The TX 0211--122 data
(G. Miley, TDT 59601713) at 160 $\mu m$ was observed in the same P32
raster mapping mode as F 10214+4724.  We followed the same procedures
for reducing the data (see \S3.4).  However, systematic effects
overwhelmed the data, and no signal was detected.  It was not possible
to carry out the specialized reduction procedures used for MG
1019+0535.  Observations by \citet{Cimatti:98} with SEST (Swedish-ESO
Submillimetre Telescope) were unable to detect the galaxy at 1270
$\mu$m ($S_{\nu}=-0.53 \pm 3.99$ mJy), but they provided a 1-$\sigma$
upper limit comparable to the 1-sigma upper limit on the detection of
MG 1019+0535 at 1250 $\mu$m.  Thus, constraints on the dust
temperature and luminosity are not as severe as those on MG
1019+0535 by this analysis.  We note, however, that of the three
galaxies discussed here, TXS 0211--122 has the least amount \lya\
quenching in relation to \heii\ or \civ\ \citep{Dey:95} (a factor of
$\sim6.5$ rather than $\sim 10$).

\section{Reconciling the FIR with Optical Data}

We now discuss the reasonableness of the upper limit temperatures and
luminosities of MG 1019+0535 and F 10214+4724 in the light of
plausible values for parameters and features characterizing their
optical spectra.

\subsection{F 10214+4724}
In the case of F 10214+4724, \citet{RowanRobinson:91} note that the
optical plus ultraviolet fluxes account for less than 1\% of the total
bolometric output of the galaxy ($\sim 0.5\%$ according to
\citet{Serjeant:98}).  This implies ${\cal L}_{FIR}\approx {\cal
L}_{bol} \approx 5.7\times 10^{46} ~\ergs \approx {\cal L}_{int}$, the
intrinsic luminosity.  This luminosity is a factor of $\sim 2$ greater
than that found by \citet{Serjeant:98} based on the $H{\alpha}$ flux
in conjunction with the magnification estimates of \citet{Downes:95}.
The observed spectral slope in UV-bright, local star forming galaxies
is found to be dependent on their color excess \citep{Kinney:93}.  If
we use this relation as a guide, then the extinction data from F
10214+4724 implies a spectral slope, defined by $F_{\lambda} \propto
\lambda^{\beta}$, of $\beta \gtrsim 0.5$, and $E_{B-V} \gtrsim 0.55$.
The slope $\beta$ is large because it has been reddened by extinction.
In this case, the intrinsic optical spectrum is energetically depleted
by $\gtrsim 99\%$.  In light of the high dust temperature found for this
object, and its starburst nature, it is reasonable to suppose that the
dust is physically close to the sources of emission.  As noted by
\citet{Chen:94}, the warm dust component surrounds OB associations and
HII regions, and attenuates the continuum, and may preferentially
attenuate the LyA line if \hone\ is present.  A cool component is
expected, but is undetected in the $\lambda \leq 1250 ~\mu$m data.  It
may be seen in the excess of the 450 and 800 $\mu$m data
\citep{RowanRobinson:93} relative to the single-temperature model
(\S3.4).

\subsection{MG 1019+0535}

What signs of optical extinction are there for MG 1019+0535?  For a
post-starburst galaxy, the OB associations and HII regions are absent.
The dust is now cool.  \citet{Dey:95} note that the decrease in the
ratio of \nv\ to \heii\ relative to the 3CR composite of
{\citet{McCarthy:93} can be explained by a large color excess
$E_{B-V}=0.3-0.4$.  But is this reasonable?  In order to calculate the
extinction in the optical, we require an estimate of the observed
continuum luminosity, a spectral slope, and a color excess.  The
former can be extracted using the flux in the redshifted line
$F_{\alpha} = 8.4 \pm 0.6 \times 10^{-17}~\ergcm2s$ and its observed
equivalent width $W_{\lambda}^{obs}=268$ \AA\ \citep{Dey:95}, leading
to a continuum flux density $F_C = 3.13 \pm 0.22 \times 10^{-19}
\ergAcm2$.  Transformed to a continuum luminosity at $z=2.765$, we
find ${\cal L}_C(\alpha) = 6.80 \pm 0.49 \times 10^{40} ~\ergs {\rm
\AA}^{-1}$.  The ``observed'' continuum depends on the spectral slope
$\beta$ and our continuum luminosity at rest $\lambda$1216 \AA .  By
increasing the slope we find that the continuum luminosity increases.
This rather counter-intuitive result follows from the fact that we are
extrapolating the continuum from the point at rest $\lambda$1216 \AA.

We calculate the extinction in optical wavelengths in order to
estimate the FIR luminosity that might result; the difference between
the unextinguished, and the observed continuum luminosity should
approximate the FIR luminosity.  The continuum optical energy lost in
extinction can be calculated by the integral, adapted from
\citet{Calzetti:94},
\begin{equation}
\Delta {\cal L}_{ext} = {\cal L}_C(\alpha) \int_{0.1
~\mu{\rm m}}^{0.6} \left\{ \left(\frac{\lambda}{0.1216 \, \mu{\rm
m}}\right)^{\beta}\, 10^{0.4 \, E_{B-V} k(\lambda)} -1 \right\} \,
d\lambda ~\ergs,
\end{equation}
where $k(\lambda)$ is a polynomial valid for Galactic extinction,
$\lambda$ is in expressed in $\mu$m, and ${\cal L}_C$ is the continuum
luminosity at the \lya\ line, given above.  If we treat MG 1019+0535
as a starburst galaxy and use the values $\beta \approx -0.5$ and
$E_{B-V}\approx 0.5$ consistent with the \citet{Kinney:93} relation
for star-forming galaxies, we can reproduce the \nv\ to \heii\ ratio
observed by \citet{Dey:95}.  However, only $\sim 2.5\%$ of the
continuum between $0.11~\mu{\rm m} \le \lambda \le 0.6 ~\mu$m would be
unextinguished.  Even so this FIR luminosity would imply a 160 $\mu$m
flux within even the 1-$\sigma$ detection limits in the ISO 160
$\mu$m-band.  However, in that case, the intrinsic \lya\ line would be
huge: To calculate the intrinsic \lya\ line luminosity, the observed
${\cal L}_{\alpha}^{obs}=4.84 \times 10^{42} ~\ergs$ must not only be
corrected by a factor of $\sim 10$ for the low line-strength of \lya\
relative to \heii , but also for the extinction at the \heii\ line,
calculated using the derivative of Eq. 1.  In this case there is a
combined factor of $\sim 130$, leading to a predicted intrinsic line
luminosity of ${\cal L}_{\alpha} = 6.3 \times 10^{44} ~\ergs$.  This
is larger than the highest \lya\ line luminosity in a sample of 165
high-redshift radio galaxies \citep{DeBreuck:00}, and roughly an order
of magnitude larger than the mean at $z \simeq 2.7$.  If we constrain
the intrinsic \lya\ line of MG 1019+0535 to be less than a more
moderate luminosity ${\cal L}_{\alpha} = 1.6 \times 10^{44} ~\ergs$ (a
value greater than $\sim 88\%$ of \lya\ line luminosities in Fig. 8 of
\citet{DeBreuck:00}), then in effect, the unextinguished, or observed
fraction of photons at rest wavelength $\sim 1640 $ \AA\ must be
$\gtrsim 1/3$.  For $\beta=0$, this requires $E_{B-V} \approx 0.12$.
For these values of $\beta$ and $E_{B-V}$ we find the energy absorbed
by dust to be a modest ${\cal L}_{ext} \simeq 4.2 \times 10^{44}
~\ergs$, and an intrinsic optical luminosity between $0.11~\mu{\rm m}
\le \lambda \le 0.6 ~\mu$m of ${\cal L}_{opt}^{int} \simeq 7.5 \times
10^{44} ~\ergs$.  These parameters would not give the observed \nv\ to
\heii\ ratio of 0.27 (it produces a ratio 0.45), but we believe that
the uncertainty in this line ratio, both in MG 1019+0535 and the
composite, is large due to the weakness of the \nv\ line.  An
intrinsic slope of $\beta \simeq 0$ is consistent with the spectral
energy distribution of a 500 Myr old starburst with solar metalicity
\citep{Leitherer:99}.

How confident are we of these values?  We have until now neglected the
effect of the uncertainties in the sub-millimeter data (see \S3.3).
We find that the maximal effect of independently varying the values
found in the sub-millimeter bands over their $\pm$1--$\sigma$ range of
uncertainties \citep{Cimatti:98}, which are on the order $\sim
20-30\%$, is slight.  For example, if $F_{160} = 2.0$ mJy, we find
that $T_d = 25.6 \pm 0.6$ K, and ${\cal L}_{FIR} = 1.43 \pm 0.20
\times 10^{46} ~\ergs$.  For $F_{160}=0.02$ mJy, $T_d = 19.2 \pm 1.2$
K, and ${\cal L}_{FIR} = 6.33 \pm 0.38 \times 10^{45} ~\ergs$, (still
about an order of magnitude greater than that which would result from
$E_{B- V}=0.12$).  Since the uncertainties apparently remain modest
and relatively uniform over the parameter range investigated here, the
uncertainties in the sub-millimeter observations can have no
significant effect on our argument that the dust cloud in MG
1019+0535 is cool.

In seeking a concordance, we have found that we must accept a larger
intrinsic \lya\ luminosity with the larger FIR luminosity when we
increase $E_{B-V}$.  To keep the intrinsic \lya\ luminosity of MG
1019+0535 from being extraordinarily large, we find we must propose a
modest color excess $E_{B-V} \approx 0.12 $.  But then we must live
without the explanation for the observed \nv\ to \heii\ ratio.

\section{Discussion}

Based on our FIR single-temperature fit, the lens-corrected bolometric
luminosity of F 10214+4724 is ${\cal L}_{tot} \gtrsim 5.7
\times 10^{46}~\ergs$.  The higher temperature of the dust, and the
highly energetic nature of the galaxy can be easily be reconciled with
a strong concurrent starburst, while the temperature of MG 1019+0535
can be reconciled with a \emph{post}-starburst galaxy.  The lack of
detection of TXS 0211--122 in the FIR is not unexpected since the
1-$\sigma$ upper limit to the flux at 1270 $\mu$m is larger than the
detection at 1250 $\mu$m for MG 1019+0535, which, after all, has a
\lya\ line more strongly diminished in relation to its \heii\ line
\citep{Dey:95}.

These galaxies appear to be detectable in the FIR in proportion to the
degree in which their \lya\ lines are surpressed (see \S1), though the
recent observations by \citet{Serjeant:98} may indicate that the \lya\
line of F10214+4724 is $\sim 4$ times stronger than that reported by
\citet{Dey:95}.  But on the other hand, the nuclear spectrum of this
Seyfert 2 galaxy may have a lower average extinction than the
continuum \citep{Serjeant:98}.  It also appears that dust temperature
may be positively correlated with the relative amount of extinction in
our sample.  Below we discuss these issues in terms of the general
properties of passively evolving starburst galaxies, and then discuss
the prospects for getting a true measurement of the dust temperature
and FIR luminosity for our two radio bright galaxies.

\subsection{General properties}

The increased frequency with which radio galaxies are found to be
observable in the $850 ~\mu$m band as one probes higher redshifts
\citep{Archibald:00} prompts us to speculate that the epoch of the
formation of the majority of radio galaxies is being probed (\ie\
$z=3-5$).  The presence of dust, albeit \emph{cool} dust, in MG
1019+0535 suggests that this radio galaxy is relatively young, yet
decidedly \emph{post}-starburst.  The starburst IRAS galaxy F
10214+4724 is found to have quite warm dust ($T_d \approx 89$ K) and
its optical luminosity is almost totally extinguished.  An interesting
``in between'' case may be the galaxy NGC 5860 (MRK 0480), which was
found to have a FIR spectrum consistent with a single temperature $T_d
\simeq 32$ K \citep{Calzetti:00}.  This galaxy has been identified as
a ``fading'' starburst \citep{Mazzarella:93}, and hence may plausibly
be construed as supporting the position that the lack of a warm
component (\ie\ $40 \lesssim T_d \lesssim 55$ K) signifies a
significant stretch of time between active star formation in the
galaxy, and the epoch of the observation.

As noted by \citet{Calzetti:00}, the dust mass involved in the cool
component of star-forming galaxies can be up to 150 times that
associated with the warm component, even though their warm component
is usually more luminous.  Thus the absence of a detected warm
component in MG 1019+0535 (or in NGC 5860) suggests that the dust mass
present in any undetected warm component is truly insignificant
compared to the mass of the cool dust.

\subsection{Modeling dust temperatures}

For a galaxy formed by an instantaneous burst, the mass ejection rate
is thought to reach a maximum dominated by supernovae within the first
$\sim 40$ Myr, followed by a steep decline of nearly two orders of
magnitude \citep{Leitherer:99}.  After this period, the mass ejection
rate is well approximated as halving as age doubles (a power law slope
of $\sim -1$).  However, the luminosity is seen to decline at a quite
similar rate.  Therefore attributing the lowered FIR luminosity for
aged radio galaxies to a dispersal of dust, rather than a decline in
galaxy luminosity, is not straightforward.  We can, however, compound
a simple model of the dust temperature as a function of distance and
time since the burst (starting, that is, at an age, $t_0 \simeq 40$
Myr).  For dust clouds well-approximated by a single temperature (as
with F 10214+4724), this seems defensible.  Further, we note that the
strong correlation of UV line luminosities and equivalent widths in
high--redshift radio galaxies implies that the source powering the
narrow line region is dominated by a common energy source
\citep{DeBreuck:00}.  We adapt the analysis of \citet{Finkbeiner:99}
regarding the equilibrium dust temperature to dust at a variable
distance $r$ from a source with a given source luminosity
characterized by a temperature $T_S$.  The dust temperature should be
well-approximated by the relation,
\begin{equation}
T_d \propto \frac{T_S^{4/(4+\epsilon)}}{r^{2/(4+\epsilon)}},
\end{equation}
where $r$ is the distance from the energy source, and the emissivity
index is assumed to be a constant $\epsilon=2$.  This model shows that
dust with $T_d \sim 20$ K must lie at a distance $\sim 80$ times
greater than dust with $T_d \sim 90$ K.  For instance, if dust at 90 K
is 5 kpc from the source, the dust at 20 K would be at 400 kpc.
However, if we accept that the luminosity of the galaxy declines
approximately with the inverse first power of time $t$ since the end
of the burst, as we noted above, but keep $T_S$ constant, then
\begin{equation}
T_d \propto (r^2 t)^{-1/(4+\epsilon)}.
\end{equation}
This equation allows us to check the hypothesis that dust is not
expelled, but that temperature differences are due to the passive
evolution of the energy source.  If the distance of the dust is
unchanged, we find that it would take $t >300$ Gyr for a dust cloud of
temperature 90 K to become a 20 K cloud under passive evolution of the
galaxy for a constant $T_S$.  Including a plausible reduction in $T_S$
to half its initial value results in $t > 18$ Gyr.  This suggests that
the principal cause of the cooler dust in MG 1019+0535 is the outward
movement of the dust cloud, rather than passive evolution of the
galaxy.  This result, it must be acknowledged, comes about by equating
the observed nature of F 10214+4724 as comparable to what MG 1019+0535
must have been like near the end of the starburst which formed it.
Though these galaxies are different in many respects, we believe they
are comparable from the standpoint of the temperature of the dust in a
passively evolving starburst galaxy.  We note that the passive
evolution of a galaxy with the apparent (lensing corrected) bolometric
luminosity of F 10214+4724 over 700 Myr could very plausibly result in
a luminosity similar to that attributed to MG 1019+0535
\citep{Leitherer:99}.

If we now allow that the cloud is being blown away at a constant
velocity $v \sim 500 ~\kms$, as suggested from the redshift of the
\lya\ line relative to \heii , \civ , and \ciii\
\citep{Dey:95,Legrand:97,Kunth:98b}, then substitution of $r=r_0+vt$
allows us to solve for the post-starburst age of the galaxy and the
galactocentric distance of the dust cloud as a function of dust
temperature.  Given a dust cloud of $\sim$90 K, assumed to be observed
at a galactocentric radius of $r_0 = 5$ kpc (roughly half of the
isophotal radius of the R-band image of MG 1019+0535 in
\citet{Dey:95}), then a $T_d=20$ K cloud could be explained if the
galaxy had evolved passively for $\sim 700$ Myr.  During this time,
the cloud would move to a distance of $r \simeq 360$ kpc.  Such a
galaxy, if seen at the redshift of MG 1019+0535 ($z=2.765$), would
have been formed at $z_ \simeq 4.3$, or 3.2, for our assumed cosmology
(\S1), or a flat universe with $h=0.7$ and $\Lambda=0.7$,
respectively.  Therefore, the naive conclusion that cool dust clouds
(in evolved galaxies lacking a warm component) are more distant than
the warm components of starburst, or fading starburst galaxies,
appears consistent with our findings, and supports the view that dust
in post-starburst galaxies is blown to large galactocentric radii by
superwinds, photon pressure, or corpuscular drag.

\subsection{Future prospects}
In order to settle the still significant uncertainties regarding the
radio galaxy MG 1019+0535 and TXS 01211-122, it would be advisable to
observe these objects near the projected maxima of their FIR spectra,
in the observed $\lambda \sim 250-500 ~\mu$m.  While SIRTF may be
useful to chart the movement of dust at redshifts out to $z \approx
1$, its wavelength coverage and angular resolution will not be
sufficient to resolve the mysteries associated with MG 1019+0535 and
TXS 0211--122.  SOFIA will have good angular resolution, but broad-band
spectral coverage stops short of the projected maximum for MG
1019+0535.  However, it may be detectable if our conclusion that $T_d
\lesssim 20$ K is incorrect.  For definitive data we will probably
have to wait until the launch of ESA's FIRST (Far InfraRed Space
Telescope), which will incorporate the bolometer instrument SPIRE,
which combines a camera and spectrometer.  It will probe point-sources
to a sensitivity $F_{\nu} \lesssim 1$ mJy in three broad bands
centered on 250, 350, and 500 $\mu$m simultaneously by the use of
dichroic filters, and have the spatial resolution needed to remove
much of the galactic cirrus.  The launch date is currently 2007.

\section{Conclusions}

We suggest the following conclusions:
\begin{itemize}
\item MG 1019+0535 is accompanied by a decidedly \emph{cool} dust
cloud, of a temperature cooler than $T_d \lesssim 20$ K, and a low
color excess in the range, $E_{B-V} \approx 0.12$, if its intrinsic
\lya\ line is not unusually luminous, but most probably no greater
than twice that.  It is likely that its bolometric luminosity is
${\cal L}_{tot} \lesssim {\rm few} \times 10^{45} ~\ergs$, consistent
with $\sim 700$ Gyr of passive evolution from a galaxy with on order
the apparent luminosity of F 10214+4724.
\item The IRAS galaxy F 10214+4724 was found to have a dust
temperature $T_d \simeq 89 \pm 12$ K, and a high FIR luminosity,
${\cal L}_{FIR} \approx 5.7 \pm 0.6 \times 10^{46} ~\ergs$, close to
its bolometric luminosity.
\item The non-detection of TXS 0211--1122 in either the 160
$\mu$m-band, or in the sub-mm at 1270 $\mu$m suggests that it has a
comparable or lower FIR luminosity and dust temperature than MG
1019+0535, in line with its 50\% larger \lya\ to \heii\ line flux
ratio.
\item The trend of high dust temperatures in starbursts, and low
temperatures in post-starburst galaxies is suggestive of dust having a
larger mean distance for evolved galaxies.  We speculate that the
principal reason for this is a combination of two factors -- first,
that dust is transported to larger galactocentric radii by various
mechanisms including photon pressure, corpuscular drag, and
superwinds, and second, the mass ejection rate ${\dot M}$
declines rapidly for a post-starburst galaxy of age $t \gtrsim 40$
Myr.  The lack of a ``warm'' dust component in a post-starburst galaxy
is then due to the steep decline in mass ejection rate at $\sim 40$
Myr post-starburst.
\end{itemize}

\acknowledgments{We wish to thank Nanyao Lu and the staff at IPAC for
hospitality and assistance during our stay in Pasadena.  We benefitted
from helpful suggestions from Arjun Dey, Douglas Finkbeiner, Daniel
Stern, and Wil Van Breugel.  We received very helpful suggestions from
the anonymous referee.  We are grateful to NASA/ISO/JPL for
financial support (contract \#961501).

\end{document}